\documentclass[fleqn,10pt, twocolumn]{wlscirep}

\usepackage{upgreek}
\usepackage{soul}
\title{Invariance properties of bacterial random walks in complex structures}

\author[1,2]{Giacomo Frangipane}
\author[1]{Gaszton Vizsnyiczai}
\author[2]{Claudio Maggi}
\author[3,4]{Romolo Savo}
\author[1]{Alfredo Sciortino}
\author[3]{Sylvain Gigan}
\author[1,2,*]{Roberto Di Leonardo}

\affil[1]{Dipartimento di Fisica, Sapienza - Universit\`{a} di Roma, Roma, I-00185, Italy}
\affil[2]{NANOTEC-CNR, Institute of Nanotechnology, Soft and Living Matter Laboratory, Roma, Italy}
\affil[3]{Laboratoire Kastler Brossel, École Normale Supérieure-Paris Sciences et Lettres (PSL) Research University, CNRS, Université Pierre et Marie Curie--Sorbonne Universités, Collège de France, 75005 Paris, France}
\affil[4]{Optical Nanomaterial Group, Institute for Quantum Electronics, Department of Physics, ETH Zurich, Auguste Piccard Hof 1, 8093, Zurich, Switzerland}
\affil[*]{roberto.dileonardo@uniroma1.it}
\begin{document}
%\keywords{Keyword1, Keyword2, Keyword3}

\begin{abstract}  
\end{abstract}
\maketitle
%\flushbottom
%\thispagestyle{empty}
%\noindent Please note: Abbreviations should be introduced at the first mention in the main text – no abbreviations lists. Suggested structure of main text (not enforced) is provided below.
%\section{Introduction}

\textbf{
Motile cells often explore natural environments characterized by a high degree of structural complexity~\cite{soil2, pseudo}. Moreover cell motility is also intrinsically noisy due to spontaneous random reorientation and speed fluctuations~\cite{bergcoli}. This interplay of internal and external noise sources gives rise to a complex dynamical behavior that can be strongly sensitive to details~\cite{reichhardt2017ratchet,koumakis2014directed, kantsler2013ciliary} and hard to model quantitatively. 
In striking contrast to this general picture we show that the mean residence time of swimming bacteria inside artificial complex microstructures, 
can be quantitatively predicted by a generalization of a recently discovered invariance property of random walks~\cite{Blanco2003}. 
We find that variations in geometry and structural disorder have a dramatic effect on the distributions of path length while mean values are strictly constrained by the sole free volume to surface ratio. 
Biological implications include the possibility of predicting and controlling the colonization of complex natural environments using only geometric informations.
}

Microfabrication provides an ideal tool to investigate the dynamics of active particles in artificial complex environments having a precise and tunable internal structure~\cite{Bechinger2016}. Arrays of scattering obstacles have been used to demonstrate the possibility of rectification and sorting in self-propelled systems~\cite{Galajda2007,hulme2008using}. 
In all these cases small details in dynamical behavior can have significant quantitative consequences. For example, the timescale of spontaneous reorientation or the distribution of scattering angles produced by obstacles have a dramatic effect on the rectification efficiency of microstructures ~\cite{reichhardt2017ratchet ,koumakis2014directed, kantsler2013ciliary }.
In striking contrast, a recently discovered invariance property of random walks implies that the average path length inside a closed domain of arbitrary shape is only proportional to the volume to surface ratio with a numerical prefactor that solely depends on the spatial dimensions.
Despite its wide generality, this invariance property has started to find applications only very recently, mainly in the field of light propagation in turbid media~\cite{pierrat2014invariance,Savo2017}.
Here we study motile bacteria exploring artificial microstructures with arbitrary geometries and a wide range of obstacle densities. Using the invariance property, we quantitatively predict the mean length of the paths traced by bacteria inside these microstructures. Assuming that the mean swimming speeds are not significantly perturbed by the presence of obstacles, we further predict and experimentally verify the invariance of the mean residence time, a more biologically relevant quantity. As a result, for all obstacle densities, the mean residence time is only determined by the microstructure volume to surface ratio and the mean inverse speed. 

We use direct laser writing by two-photon polymerization~\cite{maruo1997three,kawata2001finer} to build 3D microchambers consisting of a thick ``roof'' supported by randomly distributed pillars fabricated on a microscope coverslip. 
A 1.4~{$\upmu$}m-thick gap between the coverslip and the roof confines bacteria in a quasi-2D geometry. The entire path traced by the cells inside the microchambers is therefore restricted on the focal plane and can be easily tracked by digital video microscopy (see Supplementary Video 1). Moreover, the presence of two close surfaces prevents circular swimming caused by hydrodynamic coupling to a single solid wall~\cite{Berg1990, prx}. In all experiments we use a smooth-swimming strain of \textit{E.~coli} bacteria expressing a red fluorescent protein. Microstructures are immersed in a low-density cell suspension ($\sim 3\times10^6$~cells~$\mathrm{ml}^{-1}$, see Methods) that we track  by epifluorescence videomicroscopy. The pillars play the role of obstacles and by varying their number we can tune the degree of internal complexity of the structures. The obstacles are fabricated as randomly distributed and non-overlapping short segments measuring $3.7~\upmu$m in length and with a thickness of $0.7~\upmu$m (see Supplementary Figure 1). 
Obstacles with such an elongated shape significantly deflect colliding bacteria and  can be densely packed inside microstructures.

\begin{figure*}[ht]
\centering
\includegraphics[width=0.85\linewidth]{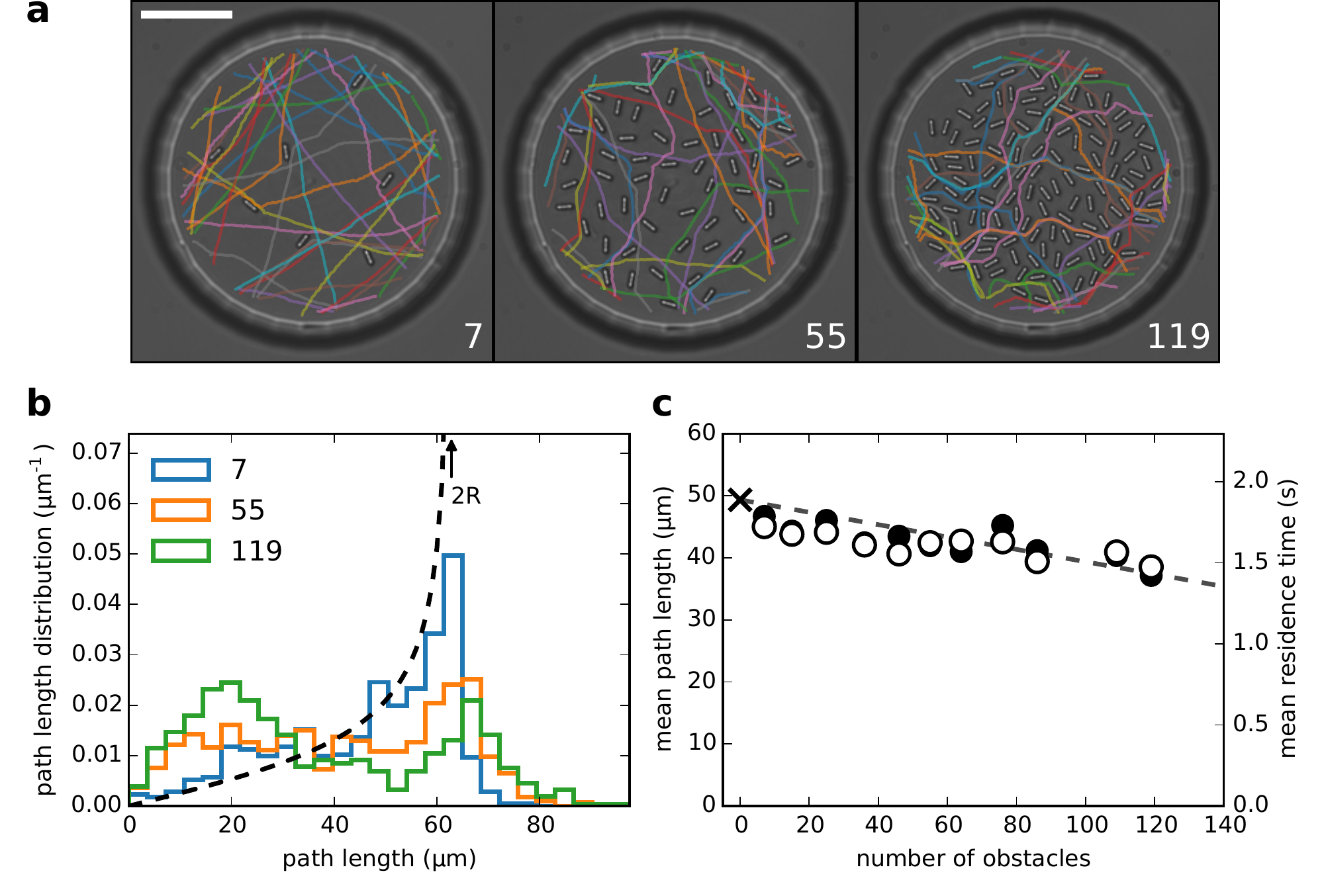}
\caption{\textbf{Invariance to obstacle density.} \textbf{a}, Optical microscopy images of three sample microstructures with number of obstacles 7, 55 and 119. The scalebar is 20~{$\upmu$}m. The colored lines are sample trajectories of swimming bacteria inside the micro-structures. As the number of obstacles increases, trajectories become more irregular due to scattering by obstacles. \textbf{b}, Path length distributions (colored bar plot) for the structures displayed in \textbf{a}. The dashed line is the theoretical prediction in the absence of obstacles. \textbf{c}, Full symbols (left $y$-axis) and open symbols (right $y$-axis) represent respectively the mean path length and the mean residence time as a function of the number of obstacles. The cross is the prediction of the BFT in absence of obstacles. The dashed line is a linear fit with the BFT formula modified to account for the excluded surface due to obstacles \ref{eqL}.}
\label{fig:2_density}
\end{figure*}
\begin{figure*}[ht]
\centering
\includegraphics[width=\linewidth]{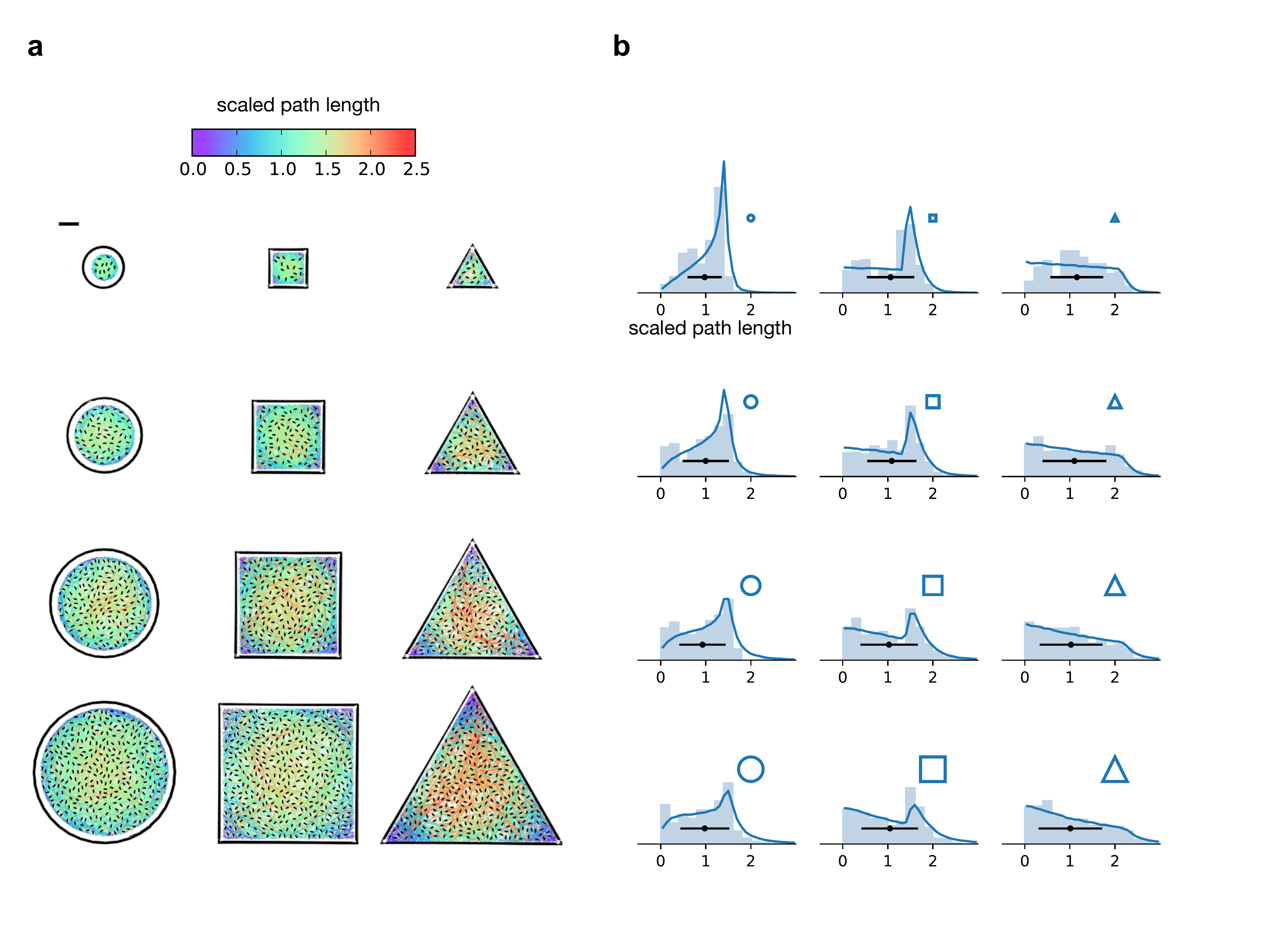}
\caption{\textbf{Size and shape effects on path length distributions.} \textbf{a}, 
We fabricate microstructures with circular, squared and triangular shapes all in four different sizes varying by an overall factor of 4 (the scalebar is 20 $\upmu$m). The density of obstacles is the same for all structures. Bacterial trajectories are plotted over microstructure images that have been processed for  contrast enhancement. Trajectories are color coded according to their scaled length (see colorbar).
\textbf{b}, Bar plots are the experimental distributions of scaled path lengths for each of the structures in \textbf{a}. The black points represent the mean scaled path length and the errorbar the corresponding standard deviation. Lines are theoretical distributions obtained by the numerical simulation of a Lorentz gas model with anisotropic obstacles mimicking experimental conditions.
}
\label{fig:3_histgrid}
\end{figure*}
We fabricate a total number of 11 circular structures with a radius $R=31~\upmu$m and with a varying number of obstacles $N=7,15,25,36,46,55,64,76,86,109,119$.
When $N$ is small, bacteria traverse the structures with approximately straight trajectories (see Fig.~\ref{fig:2_density}(a)). Since the flux of bacteria through a line element of the boundary is proportional to the cosine of the angle between the cell swimming direction and the local boundary normal, the distribution of path lengths $L$ for straight trajectories is given by the chord-length distribution ${p(L)=L/(2 R \sqrt{4 R^2-L^2})}$. As shown in Fig.~\ref{fig:2_density}(b), this distribution increases monotonically diverging at the maximum value $L=2R$. The experimental distribution of path lengths, obtained by tracking approximatively 1000 bacteria in a structure with $N=7$ obstacles, closely follows the theoretical distribution with small deviations due to rotational diffusion of bacteria and to scattering by the few isolated obstacles.
When the number of obstacles is increased, bacteria are more likely to be back-scattered by obstacles close to the domain boundary giving rise to an excess of short paths that appears as a growing peak at small path lengths.
At the same time, dense obstacles produce multiple deflections that result in trajectories with a total length that can exceed the circle diameter producing a tail in the distribution at $L>2R$. 

Despite the strong dependence on $N$ of the path length distribution, the invariance theorem by Blanco and Fournier~\cite{Blanco2003}~(BFT) can be used to obtain a strikingly simple formula for the mean path length that quantitatively matches observed values for all $N$. To this aim we now illustrate an alternative derivation of the BFT result by means of a very simple argument. 
Let's imagine to observe for a total time $T$ the motion of $N$ non-interacting random walkers moving with constant speed $v$ over a region of space of total area $\Sigma$. If reorientation dynamics is isotropic, the walkers will fill the region  with a uniform concentration $\rho=N/\Sigma$ and an everywhere isotropic velocity distribution. 
The time $t$ spent by each walker inside a generic subregion of area $S$ will be given by: 
\begin{equation}
t=T\;\frac{S}{\Sigma}
\label{t1}
\end{equation}
Alternatively we can obtain $t$ as the mean residence time $\tau$ multiplied by the number of times each walker is expected to enter the subregion during the observation time $T$. 
This number can be obtained by counting the number of bacteria that enter the subregion in a total time $T$ and then divide by $N$. Since, for an isotropic velocity distribution, the inward flux of walkers is $\rho v/\pi$ over the entire perimeter $P$ of the sub-region, we have:
\begin{equation}
t=\tau\; \frac{\rho v}{\pi}\;P\;T \frac{1}{N}
\label{t2}
\end{equation}
Equating (\ref{t1}) and (\ref{t2}) we obtain:
\begin{equation} \label{bft}
\ell=v\tau=\pi\frac{S}{P}
\end{equation}
When the subregion contains rigid obstacles, the above derivation remains valid 
as long as their presence does not significantly perturb the homogeneity of the density field in the accessible space and the isotropy of random walks along the domain boundary~\cite{Benichou2005}. This condition is fulfilled for small and convex obstacles producing scattering events lasting for a short interaction time. In this regime, however, the total area $S$ in Eq.~(\ref{bft}) must be replaced with the total accessible area $S^\prime=S-Ns$ where $s$ is the excluded area due to a single obstacle and  $N s$ the total excluded area. 
Therefore the mean path length $\ell$ is expected to decrease linearly with the number of obstacles according to the law:

\begin{equation}
\ell = \pi\frac{S}{P}\left(1-\frac{N s}{S}\right)
\label{eqL}
\end{equation}
In Fig.~\ref{fig:2_density}(c) we report as solid circles the experimental values of the mean path lengths as a function of the number of obstacles. Experimental data closely match the predicted linear behaviour in (Eq.~\ref{eqL}) with an intercept fixed at $\pi S/P=\pi R/2=49.3\;\upmu$m and a  value for ${s=6.3~\upmu\mathrm{m}^2}$
%. This value is compatible with $6.3~\upmu\mathrm{m}^2$ 
that is independently estimated by dilating the physical boundary of the obstacle by half the cell  width ($0.4~\upmu$m~\cite{schwarz2016escherichia}) (see Supplementary Figure 1). 

There are many situations in which the residence time is more relevant then the path length. For example, if adhesion to surfaces is an activated process~\cite{vissers2018bacteria} the probability that a cell will remain stuck inside a structure will be proportional to the total time spent inside it. The population average residence time can be related to the mean path length under the reasonable assumption that cell speeds $v$ are uncorrelated to path lengths $L$ so that:
\begin{equation} \label{meant}
\tau = \langle L/v\rangle=\langle L\rangle \langle v^{-1}\rangle = \ell/\bar{v}
\end{equation}
with $\bar{v}=1/\langle v^{-1}\rangle$. Although speeds can be broadly distributed in a cell population we can again use the invariance property to quantitatively predict mean residence times once the characteristic speed $\bar{v}$ is known. 
%\hl{AS: Because as long as isotropy is maintained the information about the cell reorientation is irrelevant, this allows to measure $v$ in free space in order to make predictions about both the mean path length and the mean residence time in a region in which the angular dynamics is unknown e.g. because of disorder. As an example, ... the} 
The measured mean residence time as a function of $N$ is plotted in Fig.~\ref{fig:2_density}(c) showing again a close agreement with the theoretical prediction in Eq.~(\ref{meant}) where the parameter $\bar{v}$ has been independently measured from experimental trajectories. 
%The same fitting line interpolates well the experimental values of $\langle T \rangle$ after multiplication by $\langle V^{-1} \rangle$ (see Fig.~\ref{fig:2_density}(c)).

\begin{figure}[ht]
\centering
\includegraphics[width=.965\linewidth]{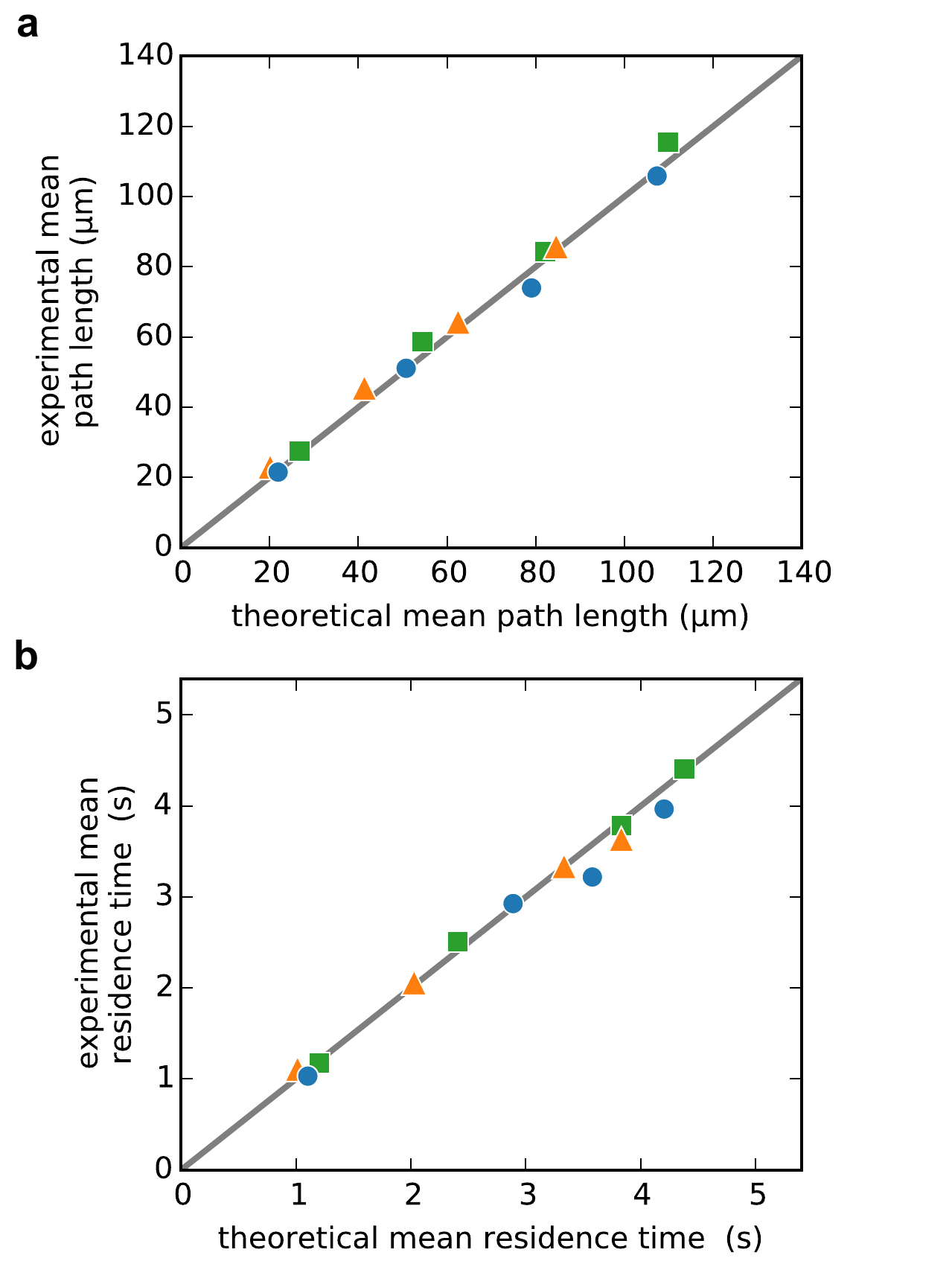}
\caption{\textbf{Size and shape invariance.} \textbf{a},
Experimental vs theoretical mean path lengths. Each shape is represented by the corresponding symbol (circle, square, triangle) appearing four times, one for each of the four sizes in Fig.\ref{fig:3_histgrid}. All structures contains obstacles with the same density. The line passing through zero represents perfect agreement between theory and experiments.
\textbf{b},
Same as \textbf{a} for the mean residence time.}
\label{fig:4_hist}
\end{figure}
We have demonstrated that, for a fixed shape, the density of obstacles has a strong effect on the distribution of path lengths while leaving the mean value unchanged as long as the total internal accessible area remains the same.
%reduces both path length and residence time by subtracting free surface from the domain. Now we turn to the question of what is the role played by the domain shape and size. 
%We demonstrated that, for a fixed shape, the number of obstacles linearly reduces both path length and residence time by subtracting free surface from the domain. 
We will now investigate the effects of shape and size on the path length distributions and again conclude that, although path distributions are very sensitive to the detailed geometry of the domain, mean path lengths and residence times can be accurately predicted using only the surface to perimeter ratio. We fabricate microstructures of circular, squared and triangular shapes all in four different sizes as shown in Fig.~\ref{fig:3_histgrid}(a). In all structures the obstacle density is fixed to ${N/S = 0.016~\upmu\mathrm{m}^{-2}}$. All recorded trajectories are also plotted in Fig.~\ref{fig:3_histgrid}(a) with a color map encoding the scaled path length $L/\ell$. %It is clear that the path lengths distribute in space more heterogeneously as the size of the structure increases, short trajectories concentrate near the boundary while long ones fill the center of the domain.
In Fig.~\ref{fig:3_histgrid}(b) we plot for each structure the histogram of path lengths normalized by the expected value $\ell$ in Eq.~(\ref{eqL}). As we move from circles to squares and then to triangles we observe that the corresponding histograms are characterized by larger frequencies for both short and long path lengths. The probability of short paths increases as we introduce convex corners and reduce their angle, as evidenced by the blue-violet color of the corners in squared and triangular structures. The presence of longer paths in triangles can be also understood noting that the normalized maximum straight paths for circles, square and triangles are respectively $4/\pi, 4\sqrt{2}/\pi$ and $4\sqrt{3}/\pi $ (see Supplementary Figure 2). Despite these qualitative and quantitative differences between histograms, mean path lengths are always close to 1 when normalized by the predicted value $\ell$  as shown in Fig~\ref{fig:3_histgrid}(b).   
%By looking at the histograms (Fig.~\ref{fig:3_histgrid}(b)) we see that, in general, the scaled path length distribution $P(\tilde{L})$ becomes broader as the size of the structure increases and, in the case of the square and the circle, the pdf develops a two peak structure. 
The observed path length distributions can be very well reproduced by a Lorentz gas model where collisions with obstacles are described as instantaneous reorientation events~\cite{lorentz}. These events are assumed to occur with a constant probability per unit path length given by 
%In this model we consider point particles moving straight at constant speed with collision happening at random (uniformly distributed) positions in space and with mean free path 
$\rho \, \sigma$, where\ $\sigma=2 b/\pi$ is the rotational average cross section of line obstacles with length $b$ (see Methods).
For perfectly straight trajectories, circles and squares have path length distributions with a vertical asymptote. A strong peak is found at the location of this divergence and is progressively smeared as the linear size of the domain becomes larger than the mean free path ${\lambda = (\rho \sigma )^{-1}}$. Upon increasing the size of the region a peak around $L=0$ appears due to bacteria that scatter close to the boundary, reorient and exit. The mix of these two types of trajectories gives rise to the observed two-peak structure. Differently for the triangle no asymptote is present in the $P(L)$ for large $\lambda$, therefore only a peak at $L=0$ progressively appears.
These results show that $P(L)$ depends strongly on the random walk properties (as the mean free path), and on the domain features (shape and size). Differently the mean values $\ell$ and $\tau$ depend only on the surface to perimeter ratio of the domain as expressed by Eq.~(\ref{eqL}). 
%The experimental values of $\ell$  are shown Fig.~\ref{fig:4_hist}(a) for all shapes and sizes. The data show a very good agreement  with the theory when we use the value ${s=6.2~\upmu\mathrm{m}^2}$ obtained above by fitting the experimental data for a varying number of obstacles (see Fig.~\ref{fig:2_density}(c)).  This value of $s$ has also been used to calculate $\tau$ according to Eq.~(\ref{meant}) which captures quantitatively well the experimental data as shown in Fig.~\ref{fig:4_hist}(b).
Experimental values of $\ell$ are shown Fig.~\ref{fig:4_hist}(a) for all shapes and sizes. We find an excellent agreement with theory using the same value ${s=6.3~\upmu\mathrm{m}^2}$ that was used previously to predict the experimental values of $\ell$ as a function of the number of obstacles (see Fig.~\ref{fig:2_density}(c)). This value of $s$ has also been used to calculate $\tau$ according to Eq.~(\ref{meant}) which captures quantitatively well the experimental data as shown in Fig.~\ref{fig:4_hist}(b).

In conclusion, we have shown that bacterial random walks inside complex and crowded microstructures do obey a general invariance property that constraints the mean value of the path length to a simple and predictable value uniquely determined by the volume to surface ratio of the structure.
More interestingly the mean residence time is also invariant as long as the volume to surface ratio remains constant. In a counterintuitive way, a greater number of obstacles does not increase the average residence time, but shortens it by decreasing the total accessible volume. The robustness of our results opens the way to different application perspectives as for example using bacteria to gather information about the internal structure of natural microenvironments. Alternatively, any significant deviation of measured residence times from the predicted value may signal the presence of physical ({\it e.g.} adhesion) and biological ({\it e.g.} taxis) mechanisms breaking the uniformity and isotropy of random walks.   

%\hl{Our observations extend the application of this invariance theorem  to the spatial and temporal dynamics of motile cells in crowded environments. The path-length and the residence -time invariance can be diagnostic tools for the bacteria colony itself and for its dynamics inside a "black-box". Specifically, deviations of mean path length from the predicted one would spot a non-isotropic velocity distribution. Instead, for bacteria moving inside an experimentally non-accessible region, where only crossing times can  be directly measured, a deviation of the mean residence time from the predicted one would pinpoint a trajectory-velocity correlation, determinate for instance from surface-adhesion.}

\bibliography{biblio}

\section*{Methods}

\subsection*{Microfabrication}
The microchamber structures were fabricated with a custom built two-photon polymerization setup \cite{vizsnyiczai2017light} from SU-8 photoresist (MicroChem Corp). A single fabrication focus was used with 8 mW laser power and 100 $\upmu \mathrm{m \; s}^{-1}$ scanning speed. After the laser fabrication scan the SU-8 photoresist sample was baked at 100 $^{\circ}$C for 7 minutes, then developed by its standard developer solvent and finally rinsed in a 1:1 mixture of water and ethanol. Strong adhesion of the microchamber structures to the carrier coverglass was ensured by a layer of OmniCoat adhesion promoter (MicroChem Corp).

\subsection*{Microscopy}
Epifluorescence imaging were performed on an inverted optical microscope (Nikon TE-2000U) equipped with a 60$\times$ (NA=1.27) water immersion objective and a high-sensitivity CMOS camera (Hamamatsu Orca Flash 4.0). %Epifluorescence was excited by Thorlabs M565L3 LED used for illumination.

\subsection*{Cell cultures}
The \textit{E.~coli} strain used is the HCB437\cite{Wolfe1987}, a smoooth-swimmer, transformed by a plasmid expressing the red fluorescent protein mRFP1 under the control of the lacI inducible promoter 
(BioBricks, BBa\_J04450 coding device inserted in pSB1C3 plasmid backbone, \href{http://parts.igem.org/Catalogue}{iGEM Catalogue}).
Cells were grown overnight in 10~mL of LB supplemented with kanamycin (30~$\upmu\mathrm{g \; mL^{-1}}$) and chloramphenicol (20~$\upmu \, \mathrm{g \; mL^{-1}}$) in a shaking incubator at 33$^{\circ}$C and 200 rpm. In the morning 50 $\upmu$L of the saturated culture was diluted (1:100) into 5~mL of tryptone broth supplemented with antibiotics and incubated in the same condition for 4~hours (until an OD$_{590}\sim$ 0.2). The production of mRFP1 was induced by adding 1~mM IPTG, and the cultures were grown for 2 further hours until exponential phase (OD$_{590}\sim$  0.6-0.8). Cells were collected and washed three times by centrifugation (1500 rcf, 5') in motility buffer (MB: 0.1~mM EDTA, 10~mM phosphate buffer, 1~mM glucose and 0.02\%Tween20).

\subsection*{Simulations}
%The computer simulations are performed by numerically evaluating the trajectories 
%of hundreds of thousands run and tumble particles.
%Particles initial potions are randomly generated along on one side of the domain (in a single point in the case of the circle). Particles enter the domain with an initial direction forming an angle with the side normal which is distributed according to a cosine law~\cite{Blanco2003}.
We simulate bacterial trajectories as straight runs intercalated by random reorientations (run and tumble dynamics) that mimic collisions with obstacles. Run lengths are exponentially distributed with a mean free path 
${\lambda=(\rho \sigma)^{-1}}$. The obstacle density $\rho$ is fixed to the experimental value ${\rho=0.016~\upmu\mathrm{m}^{-2}}$. The obstacle cross section $\sigma= 2.9~\upmu\mathrm{m}$ is evaluated as the average cross section of a randomly oriented line with length $b=4.5~\upmu$m, i.e.  ${\sigma=\pi^{-1}\int d\theta \, b \cos(\theta) =2b/\pi}$.
Assuming that an elongated particle fully aligns with a line-shaped obstacle, collisions will produce angular deflections $\delta$ that are only determined by the obstacle orientation and restricted to the interval $[-\pi/2, \pi/2]$. Weighting each outgoing direction with the cross section of the correspondingly tilted obstacle we obtain the distribution of deflection angles $p(\delta) = \sin|\delta|$ for $|\delta| \leq \pi/2$ and zero elsewhere. We also account for a small reduction in $\delta$ that could be caused by hydrodynamic interactions, causing particles to curve around the obstacle edges \cite{sipos}, or by the fact that reorientation during collision takes a finite time \cite{prx}, while the obstacle has finite length. The total deflection angle is then set to $\delta' = \delta-\mathrm{sign}(\delta) \,\alpha$, where $\alpha$ is set to $\alpha=0.39\,
(\approx \pi/8)$ to fit the histograms in Fig.~\ref{fig:3_histgrid}(b). 

\subsection*{Code availability}
The codes that support the findings of this study are available from the corresponding authors upon reasonable request.

\section*{Acknowledgements}
The research leading to these results has received funding from the European Research Council under the European Union’s Seventh Framework Programme (FP7/ 2007–2013)/ERC Grant Agreement No. 307940. S.G. acknowledges support from Institut Universitaire de France. R.S. acknowledges funding from the European Union’s Horizon 2020 research and innovation programme under the Marie Sklodowska-Curie grant agreement No. 800487. 

%\section*{Author contributions statement}
%(TO BE DEFINED...)
%... and R.D.L. designed experiments. G.F. and G.V. performed experiments. G.V. designed and fabricated microstructures. G.F. was responsible for the transformation and growth of bacterial strains. G.F., C.M. and R.D.L. analysed the data, G.F. and C.M. performed the simulations. G.F., C.M. and R.D.L. wrote the manuscript.
\section*{Data Availability}
The data that support the plots within this paper and other findings of this study are available from the corresponding authors upon reasonable request.

\section*{Additional information}
The authors declare no competing financial interests.

The corresponding author is responsible for submitting a \href{http://www.nature.com/srep/policies/index.html#competing}{competing financial interests statement} on behalf of all authors of the paper. This statement must be included in the submitted article file.

%\begin{figure}[ht]
%\centering
%\includegraphics[width=\linewidth]{stream}
%\caption{Legend (350 words max). Example legend text.}
%\label{fig:stream}
%\end{figure}

%\begin{table}[ht]
%\centering
%\begin{tabular}{|l|l|l|}
%\hline
%Condition & n & p \\
%\hline
%A & 5 & 0.1 \\
%\hline
%B & 10 & 0.01 \\
%\hline
%\end{tabular}
%\caption{\label{tab:example}Legend (350 words max). Example legend text.}
%\end{table}

\end{document}